\newcommand{\commondocopts}{letterpaper,aps,prd,10pt,superscriptaddress,showpacs,floats,nofootinbib,twocolumn,lengthcheck}
  \documentclass[\commondocopts,pdftex]{revtex4-1}
  \usepackage{cmap}
\usepackage[latin1]{inputenc}
\usepackage[T1]{fontenc}
\usepackage{subfigure}
\usepackage{amsmath,amssymb,mathbbol}

\usepackage{graphicx,color}
\usepackage{hyperref}

\newcommand{\titlestr}{A \lsgreen{vectorial} velocity filter for ultracold neutrons\\ based on a surface-disordered mirror system}
\newcommand{\comment}[1]{}

\newcommand{\ls}{\textcolor{black}}
\newcommand{\lsgreen}{\textcolor{black}}
\usepackage[normalem]{ulem}
\newcommand{\lsdel}{\comment}

\definecolor{rltred}{rgb}{0.75,0,0}
\definecolor{rltgreen}{rgb}{0,0.5,0}
\definecolor{rltblue}{rgb}{0,0,0.75}
\definecolor{rltdblue}{rgb}{0.2,0.2,0.65}
\definecolor{rltdred}{rgb}{0.65,0.2,0.2}
\definecolor{forestgreen}{rgb}{0.13,0.54,0.13}

\hypersetup{colorlinks,%
hypertexnames=true,%
pdftitle={\titlestr},%
pdfview=FitH,%
pdfstartview=FitH,%
pdfpagemode=UseNone,%
bookmarksopen=true,%
bookmarksnumbered=true,%
pdfhighlight=/I,%
linkcolor=rltred,%
citecolor=rltgreen,%
linktocpage=true}

\newcommand{\itptuw}{Institute for Theoretical Physics, Vienna University of Technology, Wiedner Hauptstra{\ss}e 8-10, 1040 Vienna, Austria, EU}
\newcommand{\iasp}{Institute of Atomic and Subatomic Physics, Vienna University of Technology, Stadionallee 2, 1020 Vienna, Austria, EU}

\begin{document}

\title{\titlestr}

\author{L.A.~Chizhova}
\email{larisa.chizhova@tuwien.ac.at}
\affiliation{\itptuw}

\author{S.~Rotter}
\affiliation{\itptuw}

\author{T.~Jenke} 
\affiliation{\iasp}

\author{G.~Cronenberg} 
\affiliation{\iasp}

\author{P.~Geltenbort} 
\affiliation{Institut Laue-Langevin, BP 156, 6, rue Jules Horowitz, 38042 Grenoble Cedex 9, France, EU}

\author{G.~Wautischer} 
\affiliation{\iasp}

\author{H.~Filter} 
\affiliation{\iasp}

\author{H.~Abele} 
\affiliation{\iasp}

\author{J.~Burgd\"orfer}
\affiliation{\itptuw}

\date{\today}

\begin{abstract}
We perform classical 3D Monte Carlo simulations of ultracold neutrons scattering through an absorbing-reflecting mirror system in the Earth's gravitational field. We show that the underlying mixed phase space of regular skipping motion and random motion due to disorder scattering can be exploited to realize a vectorial velocity filter for ultracold neutrons. \lsdel{This proposed absorbing-reflecting mirror system allows to form beams of ultracold neutrons with low angular divergence.}\ls{The absorbing reflecting mirror system proposed allows beams of ultracold neutrons with low angular divergence to be formed.} The range of velocity components can be controlled by \ls{adjusting the} geometric parameters of the system. First experimental tests of its performance are presented. \lsdel{Envisioned future applications include an investigation of transport and scattering dynamics in confined systems following the filter.}\ls{One potential future application is the investigation of transport and scattering dynamics in confined systems downstream of the filter.}
\end{abstract}

\maketitle

\section{Introduction}
\label{sec:intro}

Neutrons are conventionally referred to as ultracold when their energies are so small that they are totally reflected from materials irrespective of the angle of incidence. \lsdel{Typically, this happens for neutron energies below a few hundred neV depending on the material and allows for storage of ultracold neutrons (UCN) for long times in material or magnetic traps}\ls{Typically, neutrons are in the ultracold regime at energies below a few hundred neV (depending on the material); this allows them to be stored for long periods in material or magnetic traps} \cite{Ignatovich90, Golub79}. 
UCN are produced in a research reactor by extracting the \lsdel{low energy}\ls{low-energy} part of the Maxwellian spectrum of cold neutrons. 
\lsdel{In the last decade, considerable progress has been made 
to experimentally realize superthermal sources of UCN.} 
\ls{Considerable progress has been made \cite{Zimmer2011,Lauer2010,Trinks2000,Masuda00,Bake03} over the last decade to produce superthermal sources of UCN experimentally \cite{Golub75}.} Well known UCN physics applications include measurements of the neutron electric dipole moment \cite{Bake03,Beringer12}\lsdel{, the} \ls{and of the} neutron half-life time \cite{Beringer12} as well as the observation of the UCN quantum states in the Earth's gravitational field \cite{Nesv02, Jenke11}. 

UCN can also serve as a powerful tool for studying quantum chaos \cite{Stockmann99}, \ls{the} scattering of ultracold fermions \cite{Schneider12, Brantut12} and \ls{the} transport \ls{of fermions} in confined geometries in the presence of only short-ranged interactions\ls{,} while eliminating distortions by Coulomb correlations and interactions. For these kind of experiments, a UCN beam with well-defined velocity and angular distribution is needed. However, in \lsdel{the experiment}\ls{experiments} without an additional selector, \lsdel{such a condition}\ls{this} is hard to achieve since the velocity spectra in the initial UCN beam are usually broad.  
In this \lsdel{paper, we investigate theoretically and experimentally} \ls{paper we assess, theoretically and experimentally,} the efficiency of a band-pass filter for the vectorial velocity of UCN to form a beam of UCN with a well-defined velocity distribution and small angular divergence. The filter consists of a neutron absorbing-reflecting mirror system (ARMS) similar to \lsdel{previously suggested setups that were used for other purposes such as}\ls{the setups already proposed and used for} the investigation of neutron quantum states \cite{Nesv02,luschikov78,Westphal07}, the preparation of wave packets to study the dynamics of the quantum bouncing ball~\cite{Jenke09} or the highly-efficient extraction of UCN from storage volumes~\cite{Schmidt07, Barnard08}. \lsgreen{In contrast to these earlier works however, we operate this system in the classical parameter regime; it thus serves as a beam shaper, thanks to its mixed regular and chaotic phase space structure with (moderately) high flux. For sampling the classical phase space we employ the classical trajectory Monte Carlo (CTMC) method which ignores all quantum effects related to diffraction, localization, correlation volumes, finite correlation length etc. \cite{Rauch00, Gahler98, Felber98}. The limits and validity of a classical description for ultracold neutrons have been extensively discussed in the literature (see for example \cite{Rauch00,Rauch96, Gahler98, Felber98}). The good agreement between our theoretical analysis and the experimental results in the regime of large mirror separation does suggest that our classical simulation provides a good approximation to the situation in the experiment. This classical scattering analysis also allows us to derive analytical estimates for the cut-offs of the components of the velocity vector after the filter.
We expect this mirror system to provide a well controlled incident flux of UCN for future quantum transport, quantum scattering and quantum chaos experiments.}

\section{Set-up of the filter}
\label{sec:setup}

\begin{figure}[t]
 \centering
 \includegraphics[width=\linewidth]{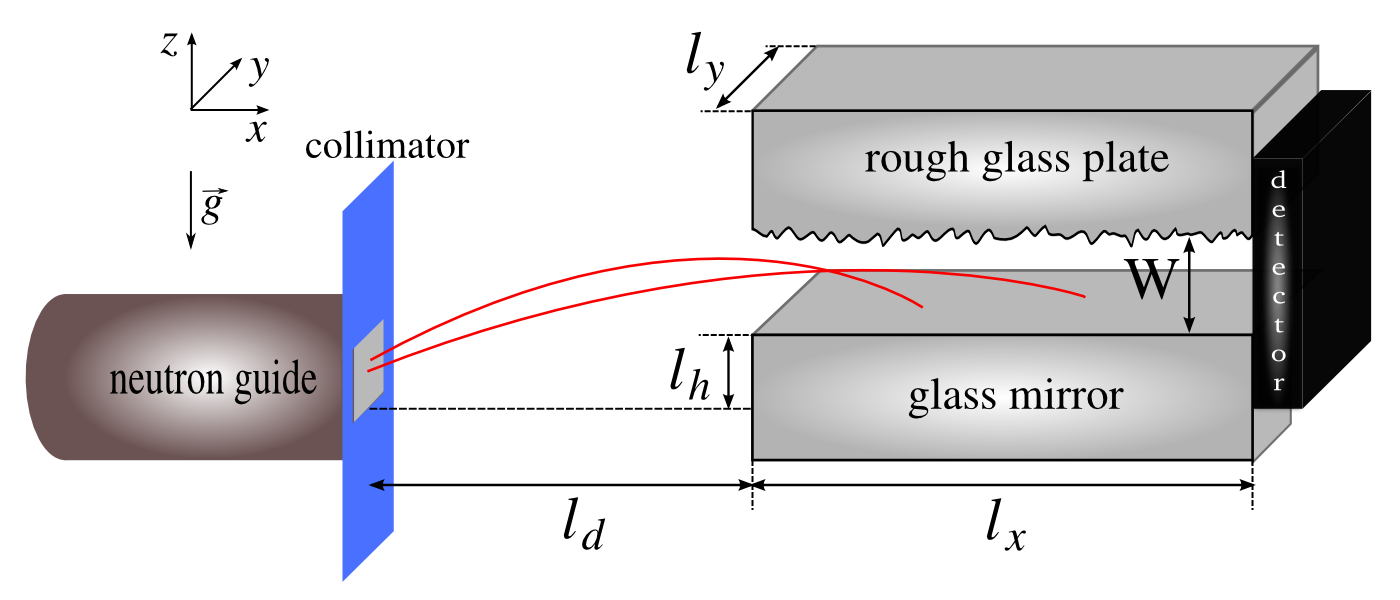}
 \caption{(Color online) Schematic view of \ls{the} ARMS proposed as a filter for the vectorial velocity $\vec{v}$\lsdel{ consisting}\ls{. The system consists} of a glass mirror (bottom) and a rough glass plate with surface disorder inducing large angle disorder scattering and absorption (top). The incident neutron beam leaves the neutron guide through a collimator (for the position of the spacers see Fig.~\ref{fig:limvel}(b) below).}
 \label{fig:system}
\end{figure}

\begin{figure}[t]
 \centering
 \includegraphics[width=0.7\linewidth]{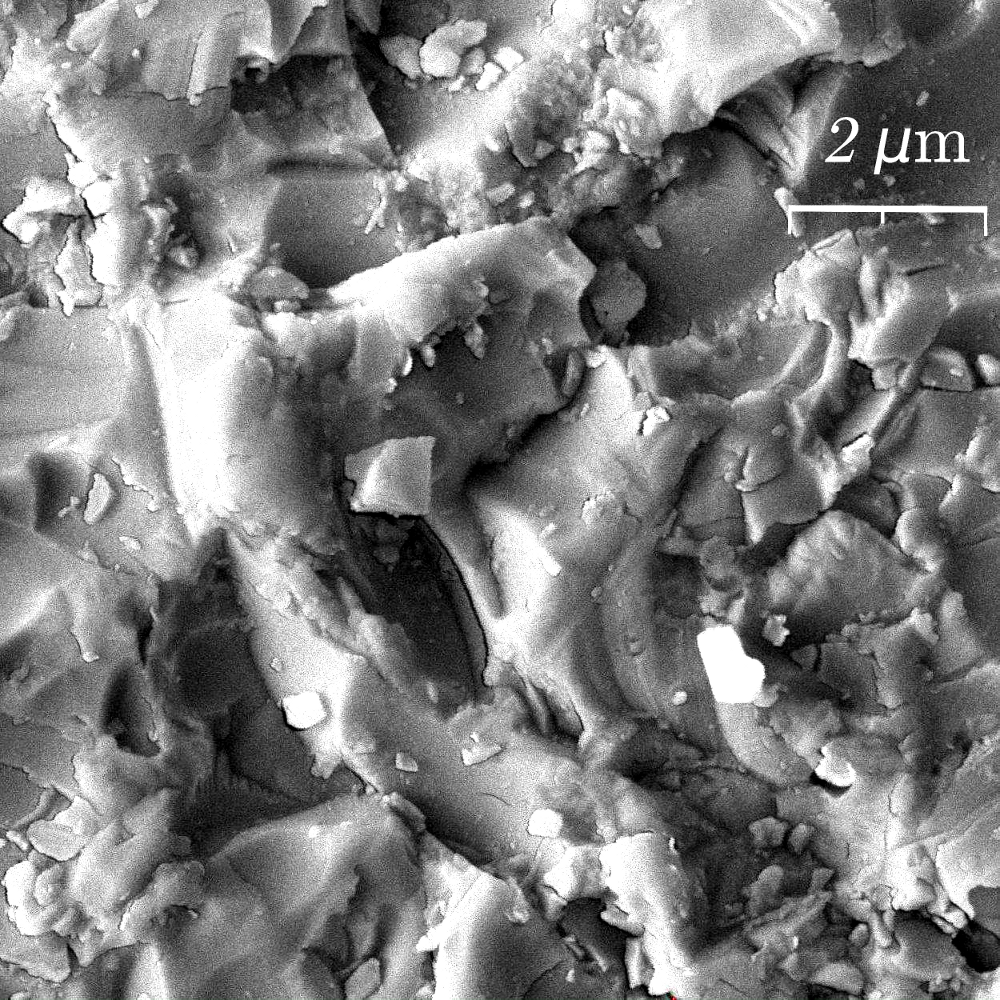}
 \caption{A typical surface of the rough upper mirror measured with the scanning electron microscope (SEM). To produce an SEM image the surface was coated with Ni making the sample electrically conductive. (The measurement was performed at the USTEM facility \cite{USTEM}.) The disorder has amplitude $A \sim 3\: \mu$m and a correlation length $l_{cor} \sim 2\: \mu$m.}
 \label{fig:profile}
\end{figure}

Figure \ref{fig:system} shows a schematic \lsdel{picture}\ls{view} of the proposed absorbing-reflecting mirror system (ARMS). The UCN beam from the neutron guide \lsdel{passing a collimator}\ls{crosses a collimator and} enters the ARMS which consists of two horizontal borosilicate glass plates, or mirrors, separated by a distance $W$. We investigate this system in the classical regime, where quantum effects such as diffraction or interference \cite{Rauch00} and localization \cite{Anderson58} can be neglected. This helps to form a beam with (moderately) high flux. Moreover, the properties of the ARMS can be described by a classical phase space analysis. The two glass plates are separated by two spacers\lsdel{which}\ls{; these} are rectangular blocks with height $W$ and \lsdel{a}length equal to the mirror length $l_x$ (see Fig.~\ref{fig:limvel}(b), not shown in Fig.~\ref{fig:system}) placed symmetrically \lsdel{at the sides}\ls{on each side} of the glass plates. For the experiment, the spacers are made of brass which is nonmagnetic and has a comparatively low critical velocity $v_{cr}$. The \lsdel{inner surface of the {\it upper} mirror is rough where the roughness is}\ls{roughness of the inner surface of the {\it upper} mirror is} characterized by the amplitude $A$ of the disorder and its correlation length $l_{cor}$. A typical surface profile measured with a scanning electron microscope is shown in Fig.~\ref{fig:profile}. The inner surface of the {\it lower} mirror is flat allowing for specular reflections. \lsdel{Only the neutrons which pass through this device without being absorbed by the glass plates (or by the spacers) contribute to the flux of the emerging beam.}\ls{The flux of the final emerging beam consists of the neutrons traversing this device without being absorbed by the glass plates (or spacers).} The \lsdel{latter}\ls{flux} can be measured by a detector \lsdel{that covers an}\ls{with a surface} area of $W \times l_y$ (Fig.~\ref{fig:system}). 

To understand how this device \lsdel{can operate}\ls{functions} as a velocity filter and \lsdel{a}beam shaper, 
we analyze the classical phase space of this system with mixed regular-chaotic dynamics. 
A neutron traveling between the two mirrors has three degrees of freedom and its Hamiltonian 
is given by
\begin{equation}
\label{eq:Ham}
H(p_x, p_y, p_z, x, y, z) = \frac{p^2_x}{2m} + \frac{p^2_y}{2m} + \frac{p^2_z}{2m} + mgz + V(x,y,z),
\end{equation}
where $g$ is the Earth's gravitational acceleration and $V(x, y, z)$ describes the potential of the walls approximated by
\begin{equation}
\label{eq:Pot}
V(x,y,z) = \begin{cases} 0, & \mbox{if } 0 < z < \xi (x,y) \\ V_F, & \mbox{otherwise, }  \end{cases}
\end{equation}
where $V_F$ is the Fermi pseudopotential to be discussed in more detail below and $\xi(x,y)$ describes the disorder landscape of the upper disordered surface. 
The Hamiltonian in Eq.~(\ref{eq:Ham}) governs the trajectories in 6-dimensional phase space. We note parenthetically that quantum simulations for the present system in its full dimensionality are \lsdel{currently}still out of reach (see, e.g., \cite{Meyerovich06, Westphal07, Voronin06}). 
For the parameter regime under investigation\ls{,} a classical description is expected to be 
valid\ls{,} as the discrete nature of the transverse energy due to the confined potential 
Eq.~(\ref{eq:Pot}) is negligible in the limit of wide openings, i.e.\ for large quantum numbers with many thousands of 
open scattering modes (for $W=200 \: \mu$m typically $N_{sc} \gtrsim 10^3$). 
Moreover, since the de Broglie wavelength of the UCN ($\lambda_D \lesssim 100nm$) is much 
smaller than the typical surface roughness ($\sim 2 \: \mu$m), quantum and classical 
dynamics are expected to closely resemble each other. 

\begin{figure}[t]
 \centering
 \includegraphics[width=0.7\linewidth]{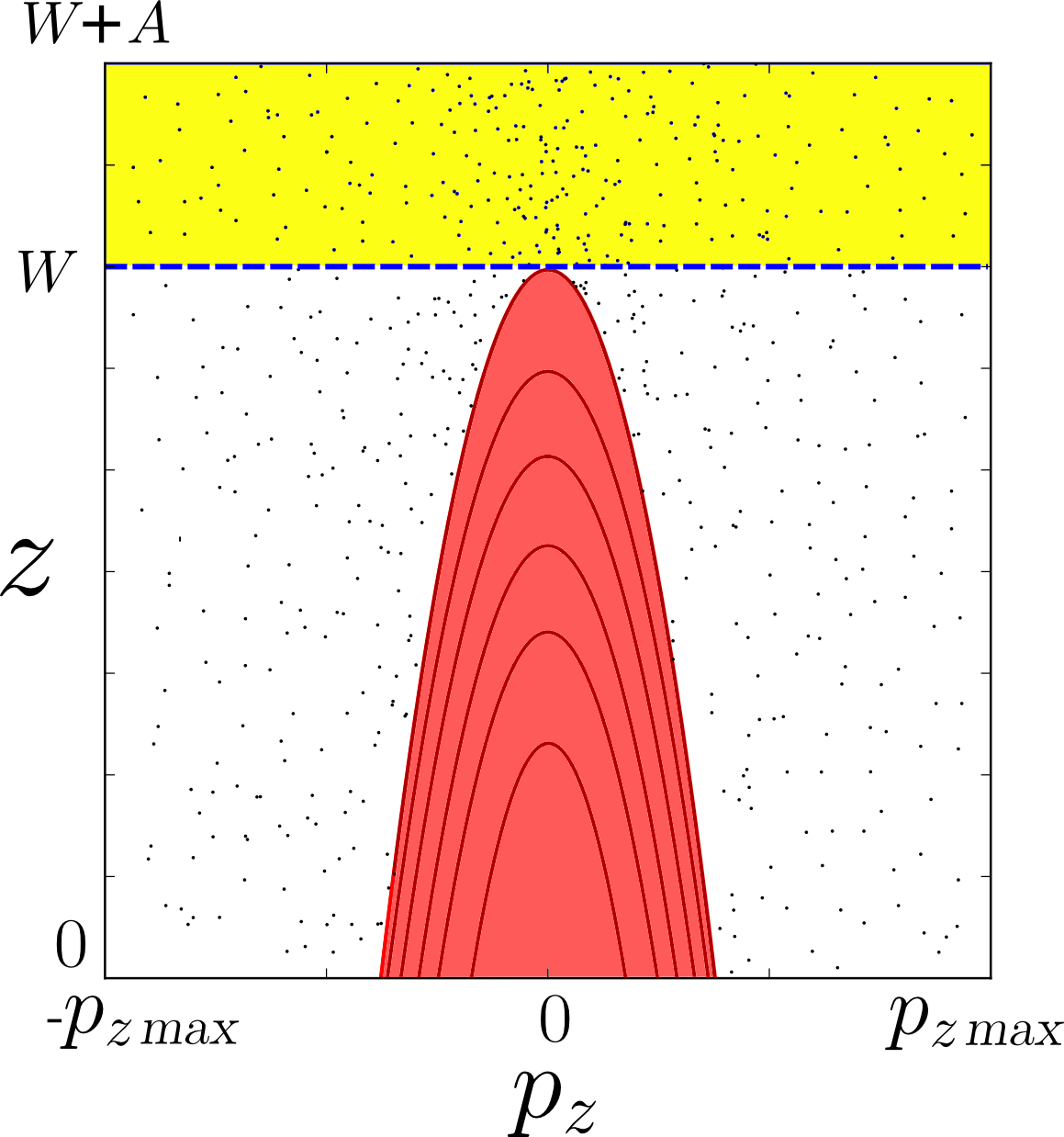}
 \caption{(Color online) The projection of the Poincar\'e surface of section onto the $(p_z, z)$ plane of the phase space for neutron transport consists of a regular island (red (gray) area at the center) representing skipping trajectories, and of an irregular part of chaotic scattering trajectories induced by the surface disorder with roughness amplitude $A$. The yellow (light gray) area with $z$ ranging from $W$ to $W + A$ indicates the region where neutrons scatter off the upper surface.}
 \label{fig:poincare}
\end{figure}

A useful method for the visualization of classical trajectories lying on the 5-dimensional 
isoenergy surface in multi-dimensional phase space is the projected Poincar\'e surface of 
section. 
\lsgreen{This method is particularly useful for answering the question whether the system is integrable or not -- also in cases like the present one, 
where the trajectories lie on different isoenergy surfaces due to the thermally distributed inital conditions of neutrons leaving the wave guide
\cite{Lichtenberg92, Reichl92}.} \lsgreen{We start from a four-dimensional surface of section given by a cut at the system 
exit $x = l_x$ and impose periodic boundary conditions in $x$ direction to visualize the 
different types of trajectories (regular vs.~chaotic) present in the system  \cite{Lichtenberg92}. (Note that 
no periodic boundary conditions are present in the real system, where the Poincar\'e surface
of section would consist only of a single intersection of each neutron trajectory at the moment the 
neutrons leave the mirror setup).} We then project the 
surface of section onto the $(p_z, z)$ plane for $p_x > 0$. Fig.~\ref{fig:poincare} clearly 
displays two types of motion (note that due to time-reversal symmetry the surface of section 
for $p_x < 0$ is equivalent). The large regular island at the center corresponds to the 
neutrons repeatedly bouncing off the lower mirror in the Earth's gravitational field 
({\it skipping} motion, \cite{Ohtsuki79}) and traveling through the filter without touching 
the \lsdel{upper rough}\ls{rough upper} wall. For this motion the degrees of freedom are separable and the 
trajectory intersection lies on a smooth curve in the $(p_z, z)$ plane. The equation 
for this curve $p_z = p_z(z)$ follows from \lsdel{the Hamilton equation with}the Hamiltonian 
$H(p_z,z)=H_{z0}$ being the integral of motion and independent of the other two degrees 
of freedom. It is explicitly given by
\begin{equation}
p_z = \pm m \sqrt{2g(h - z)},
 \end{equation}
where $h$ is the maximal height that a neutron can reach for a given energy $H_{z0} = mgh$.
For large $W$ (typically $W \gtrsim 200 \mu$m in our experiment) the regular island 
associated with skipping motion supports a large number ($N_{sk} \gtrsim 40$) of transverse 
quantum states (though still a small subset of all open modes)\ls{,} indicating that the flux 
associated with the island is in the (semi) classical regime. Since this flux scales with 
$W^{3/2}$ (the area of the island), large $W$ are advantageous for achieving a (relatively) 
high flux.   

The region of skipping motion in phase space is the complement to the region of 
{\it chaotic} motion \lsdel{of those neutrons that scatter off}\ls{of the neutrons scattering off} the disordered upper surface creating the chaotic sea in phase space. In the latter case, due to the roughness, the degrees of freedom are not separable and additional integrals of motion, besides the total energy, no longer exist. Therefore, the intersection of a randomly scattered trajectory with the projected surface of section does not form \ls{a} smooth curve \lsdel{and}\ls{but} shows chaotic behavior instead \cite{Lichtenberg92}.        

The interaction of \ls{the} neutrons with the surface of the glass walls \lsdel{as well as}\ls{and} the spacers is described by a Fermi potential $V_F$ which characterizes the low-energy nuclear scattering of surface atoms \cite{Ignatovich90, Golub79} and can be written as
\begin{equation}
\label{eq:Fermi}
V_F = V_R - iV_I = \frac{2 \pi \hbar^2}{m}N(b - ib_i), 
\end{equation}
where $m$ is the neutron mass, $N$ is the scattering center density and $b$ is the coherent nuclear scattering length. The imaginary part of the potential is determined by
\begin{equation}
\label{eq:b_i}
b_i = \frac{\sigma_l v}{4 \pi \hbar},
\end{equation}
where $v$ is the neutron velocity and $\sigma_l$ is the absorption scattering cross section due to the interaction of neutrons with nuclei of surface atoms. The absorption results from neutron capture by nuclei accompanied by emission of $\gamma$-rays or charged particles. This effect is accounted for \lsdel{with}in our description of elastic neutron transport by introducing the effective imaginary potential $V_I$ in Eq.~(\ref{eq:Fermi}). \lsdel{Within}\ls{In} the classical simulation, it gives rise to a loss probability for the trajectories \lsgreen{discussed below (see Eq.~(\ref{eq:refl1}) and Eq.~(\ref{eq:refl}))}.    

The real part $V_R$ of the Fermi potential determines the critical energy and velocity 
$v_{cr}$ of \ls{the} neutrons which are reflected. If a neutron has \lsdel{a}\ls{higher} kinetic energy in the direction 
normal to the surface \lsdel{larger}than the real part of the Fermi potential of the surface 
material ($E_{\perp} > V_R$), it penetrates the wall and is considered to be lost. 
The surface roughness gives rise to nonspecular reflection and, thus, to \ls{the} transfer of 
longitudinal ($E_{\parallel}$) to transverse ($E_{\perp}$) energy of neutrons. 
As a result, the total absorption \lsdel{will be}\ls{is} enhanced by surface disorder. 
Consequently, neutrons scattered off the upper wall, i.e., neutrons undergoing chaotic 
scattering, are eventually removed from the transmitted current. 
\lsdel{Effective}\ls{The effective} removal of neutrons with unwanted velocity components $v_{\perp} > v_{cr}$ is 
key to the beam\ls{-}shaping properties of the filter. \lsdel{Therefore materials}\ls{Materials} with (relatively) 
low $v_{cr}$ are \ls{therefore} advantageous. 

Even for $E_{\perp} < V_R$ the reflection probability from the mirror surfaces is below one due to an exponential penetration of the wave function into the barrier (tunneling). The reflection probability of the neutron from the step-like potential barrier follows as (e.g. see \cite{Golub79, Landau77})
\begin{equation}
\label{eq:refl1}
R = \left| \frac{k_0 - k}{k_0 + k} \right| ^2 = \left| \frac{E_{\perp}^{1/2} - (E_{\perp} - V_F)^{1/2}}{E_{\perp}^{1/2} + (E_{\perp} - V_F)^{1/2}} \right| ^2, 
\end{equation}
where $k_0$ and $k$ are the normal components of the wave vector in vacuum and in the medium. 
For the case $V_I \ll V_R$ (usually $V_I$ is $10^4$ times smaller than $V_R$) the reflection 
coefficient, to first order in $V_I/V_R$, becomes \ls{\cite{Golub79,ignat96}}
\begin{equation}
\label{eq:refl}
R = 1 - \mu (E_{\perp}) = 1 - 2 \frac{V_I}{V_R} \sqrt {\frac{E_{\perp}}{V_R - E_{\perp}}}
\end{equation}
with $E_{\perp} < V_R$. \lsgreen{This expression contains the important wall-loss probability per bounce, $\mu (E_{\perp})$, which drops to zero for vanishing imaginary part of the
Fermi potential. In this case the cross-section for neutron capture in the walls is zero, resulting in total reflection, $R=1$.} 
It should be noted that the loss factor $V_I/V_R = \sigma_l v/4 \pi \hbar b$ is independent 
of the neutron velocity since $\sigma_l$ is proportional to $1/v$ \cite{Ignatovich90}. 
Multiple scattering leads to preferential depletion of the chaotic region of the phase space 
since the number of bounces \lsdel{at}\ls{off the} walls for chaotic scattering trajectories is much larger than 
for skipping trajectories.

\section{Monte Carlo Simulation}
\label{sec:MCS}
To explore the implications of the mixed phase space structure analyzed above for the neutron transport through the ARMS, we perform a full three-dimensional Monte Carlo simulation using \ls{the} typical geometric parameters of the ARMS and realistic initial distributions of the neutron beam. 

\subsection{Geometry of the set-up}
\label{sec:MCS_geo}
\lsdel{The two parallel absorber-mirrors are chosen to be square plates with}\ls{Two square plates are used for the two parallel absorber-mirrors, with} $l_x = l_y = 10$ cm 
(see Fig.~\ref{fig:system}).  In order to achieve a sizeable flux and to stay within the 
classical regime for the distance between the two mirrors, a relatively large value of 
$W = 250 \: \mu$m is \lsdel{chosen}\ls{used}. The mirrors are made of borosilicate glass and have a critical 
velocity $v_{cr} = 4.3$ m/s and a loss factor (see Eq.~(\ref{eq:refl})) of 
$V_I/V_R = 0.0015$ \cite{Ignatovich90}. The brass spacers shown in Fig.~\ref{fig:limvel}(b) 
have a critical velocity $v_{cr} = 5.3$ m/s \cite{Ignatovich90}. To simulate the landscape of 
surface disorder $\xi (x,y)$ in the Hamiltonian of Eq.~(\ref{eq:Ham}) with realistic 
spatial correlations we use a convolution method with a Gaussian correlation function
\begin{align}
\label{eq:cor}
\langle \xi(x,y) \xi(x+x', & y+y') \rangle = \\ \nonumber
			& A^2 \exp \left( -\frac{x'^2}{2l_{cor}^2} \right) \exp \left(-\frac{y'^2}{2l_{cor}^2} \right).
\end{align}
The description of this method can be found in \cite{Rice54, Izrailev11}.  The amplitude and the correlation length of the surface disorder in the simulations are chosen to be $A = 3 \: \mu$m and $l_{cor} = 2 \: \mu$m\ls{,} taken from the measured surface profile (Fig.~\ref{fig:profile}). The distance between collimator and mirrors is $l_d = 18$ cm\ls{. The} \lsdel{the}distance between the lower collimator edge and the lower mirror surface is $l_h = 5.4$ mm. The size of the collimator opening is $y \times z$ = $8$ cm $\times 2.7$ mm. We emphasize that the present simulation of the two-mirror setup\ls{,} taking into account also the transverse degree of freedom $y$\ls{,} is fully three-dimensional\ls{,} unlike earlier two-dimensional models for similar systems.

\subsection{Initial conditions}
\label{sec:MCS_init}

The simulation requires the phase space distribution $\rho(\vec{r},\vec{v},t=0)$ of the UCN beam exiting the collimator as initial conditions for the propagation. The most reliable information about the distribution of the forward velocity component~$v_x$ of the neutron beam at the exit of the neutron guide can be obtained from time-of-flight (TOF) measurements. In \lsdel{such a measurement, the}\ls{these measurements the} neutron beam is periodically chopped (see, e.g.~\cite{Fierlinger06}). After this chopper, the neutrons traverse a neutron guide of known length before they are detected with the help of a neutron counter. The detector electronics also receive one logical signal per turn of the chopper blade and thereby measures the time of flight of the neutrons.

\begin{figure}[t]
 \centering
 \includegraphics[width=\linewidth]{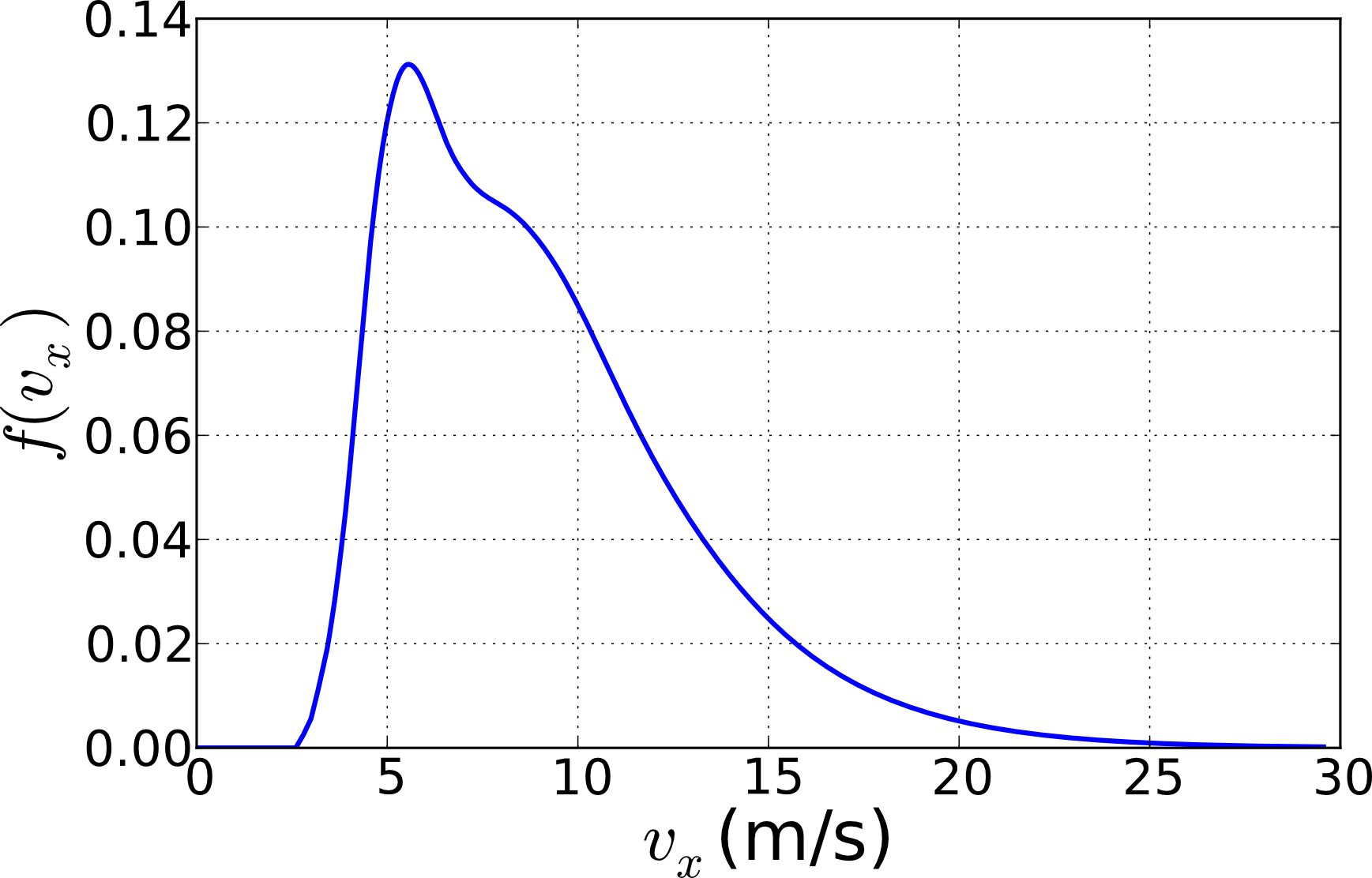}
 \caption{(Color online) Velocity distribution $f(v_{x})$ measured by time-of-flight at the ILL facility for a straight, $2$ m long cylindrical neutron guide with $4$ cm radius (beamline UCN at PF2).}
 \label{fig:tof}
\end{figure}

The time-of-flight spectrum of the ultracold neutron beam at the Institute Laue-Langevin was measured with this method \lsdel{in 2008}and corrected for the chopper offset \cite{Jenke11a}. The opening function of the chopper was deconvoluted and with the help of the distance-of-flight, the corresponding velocity spectrum was extracted.
The resulting normalized distribution of the forward velocity in $x$-direction (Fig.~\ref{fig:tof}) will be used as \ls{the} initial distribution function for $v_x$. 

Although it is difficult to measure the transverse velocity components $v_y$ and $v_z$, partial information on the angular distribution $P(\theta)$ of the beam are available (see, e.g.~\cite{Ignatovich90, Kosvintsev77}). $P(\theta)$ depends on the ratio $R$ of length to radius of the neutron guide. Since the angular distribution for the pipe which was used for the TOF measurement is unknown we perform our calculations for two distributions which correspond to the following limiting cases: in \ls{the} case of the ratio $R \gg 1$ the angular distribution is well approximated by $P(\theta) \sim \cos(\theta)$  and in \ls{the} case of a short pipe with $R \gtrsim 1$ the angular distribution is uniform. Because of the cylindrical symmetry of the pipe and the small angles involved, we consider the distributions $P(\theta_y)$ and $P(\theta_z)$ to be identical and either cosine or uniformly distributed. We characterize in the following the vectorial velocity distribution of the neutron flux through the ARMS by the Cartesian angles,
\begin{subequations}
\label{eq:angles}
   \begin{equation}
      \theta_y = \tan^{-1}(v_y/v_x),
   \end{equation}
   \begin{equation}
      \theta_z = \tan^{-1}(v_z/v_x).
   \end{equation}
\end{subequations}
Controlling and reducing the divergence, in particular in $\theta_z$, of the UCN flux is \lsdel{one}\ls{a} key feature of the ARMS \lsdel{that provides}\ls{, providing} a well focused beam for quantum transport studies in the gravitational field. Since the inner surface of the cylindrical neutron guide is coated with $^{58}$Ni, which has a critical velocity of $8.07 \: m/s$, the velocity component normal to the guide surface at the pipe exit $v_{\perp} = \sqrt{v_y^2 + v_z^2}$ \lsdel{should be}\ls{must be} less than $8.07 \: m/s$. The initial vectorial velocity distribution is randomly sampled from these distributions. The coordinate distribution of the neutron beam in the plane of the collimator opening is assumed to be uniform.

The classical equations of motion are integrated for a large number of initial conditions $
\rho(t=0)$. Calculations start at the collimator opening with initial conditions randomly 
selected for each particle from the distributions discussed in this subsection. 
The number of neutrons that reach the mirror slit entrance is kept fixed at $5 \cdot 10^5$ 
for the results presented in this section \lsdel{so that the statistical errors are kept small}\ls{to reduce statistical errors}. The propagation inside the 
slit includes absorption \lsdel{by walls and by spacers}\ls{by the walls and spacers} when the kinetic energy perpendicular 
to the surface is larger than the Fermi potential. The wall-loss probability per 
bounce Eq.~(\ref{eq:refl}) is included as a stochastic process for each collision 
with the wall. 

\subsection{Simulation results}
\label{sec:MCS_sim}

\begin{figure}[t]
 \centering
 \includegraphics[width=\linewidth]{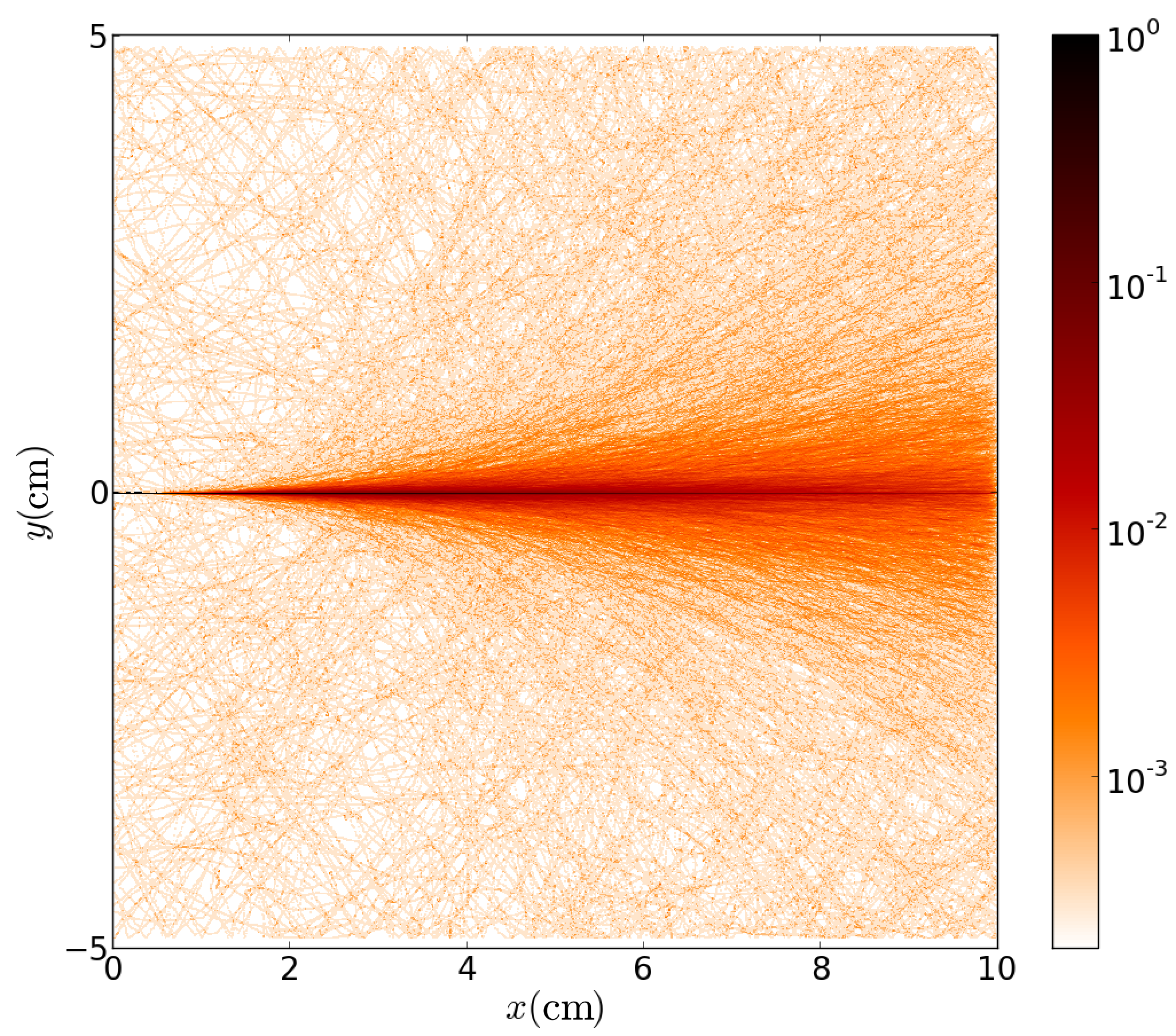}
 \caption{(Color online) A sample distribution of 3000 trajectories undergoing chaotic scattering in the top view of the absorbing-reflecting mirror system (ARMS). The trajectories enter with initial conditions $y = 0$ and $v_y = 0$. Color scale in arbitrary units.}
 \label{fig:traj}
\end{figure}

An example of the trajectory distribution between \ls{the} two mirror plates \lsdel{in top view}\ls{viewed from the top} of the system is shown in Fig.~\ref{fig:traj}. The trajectories depicted here are calculated for 3000 different initial conditions lying inside the region of chaotic scattering. For better \lsdel{visualization} \ls{visibility} we \ls{have} restricted \lsdel{here}the initial conditions to $y = 0$ and $v_y = 0$. \lsdel{We also do not show the skipping trajectories as they would coalesce to a single line}\ls{For the same reason we do not show the skipping trajectories, as they would have coalesced into a single line} ($y=0$, in this case) and would \lsdel{decrease}\ls{reduce} the contrast of the irregular scattering pattern. 

\begin{figure*}[p]
 \centering
 \includegraphics[width=0.9\textwidth]{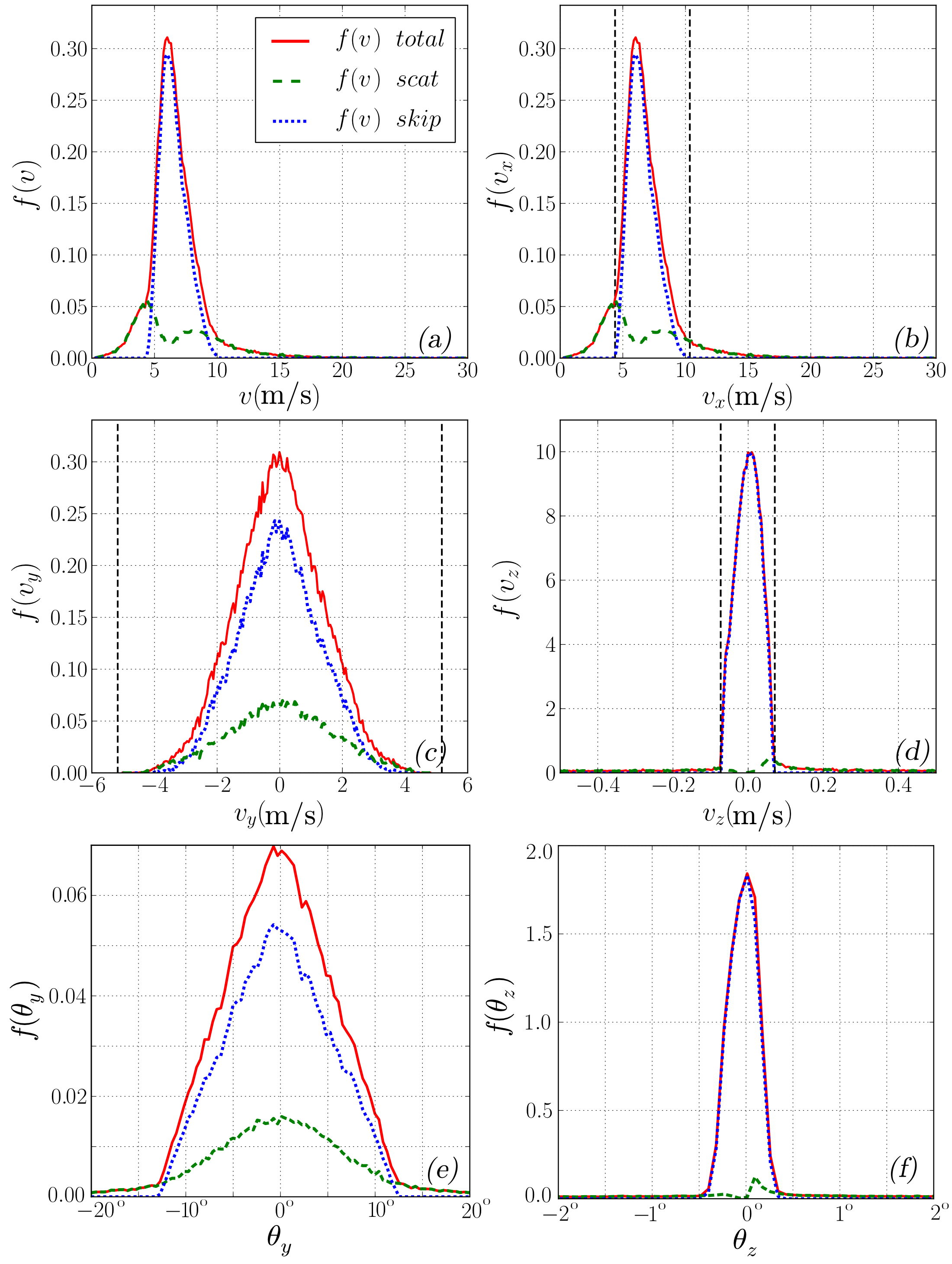}
 \caption{(Color online) Monte Carlo simulations for the components of the vectorial velocity distribution after passing the ARMS with parameters discussed in section \ref{sec:MCS_geo}. (a) Distribution of the absolute value of the total velocity $f(v)$; (b-d) distributions of the components $f(v_x)$, $f(v_y)$ and $f(v_z)$; (e, f) angular distributions $f(\theta_y)$ and $f(\theta_z)$. The total distributions (solid red lines) can be decomposed into gravitational skipping (dotted blue line) and chaotic scattering (dashed green line) trajectories. Vertical lines correspond to analytical results for the band-pass of the velocity components Eq.(\ref{eq:limitvel}-\ref{eq:v_ylim1}). The parameters of the filter: $W = 250\: \mu$m, $l_x = l_y = 10$ cm, $A = 3\: \mu$m, $l_{cor} = 2\: \mu$m.}
 \label{fig:resultsp_cos}
\end{figure*}

Typical results of the classical Monte Carlo simulations of the full 3D filter system are presented in Fig.~\ref{fig:resultsp_cos}, where distributions $ f(v_i)$ of the absolute value of the total velocity (a) as well as the $v_x$, $v_y$ and $v_z$ velocity components (c-d) at the detector position are shown for an initial cosine angular distribution. If a uniform initial angular distribution in the beam is used instead, the results hardly differ from the ones presented in Fig.~\ref{fig:resultsp_cos}. This is a consequence of the fact that the initial \lsdel{maximal}\ls{maximum} angles are relatively small ($\theta_y \sim 26^{\circ}$ and $\theta_z \sim 7^{\circ}$). Therefore, the uniform and cosine distributions do not significantly differ from each other. The angular distributions $f(\theta_y)$ and $f(\theta_z)$ after passing the ARMS are presented in Fig.~\ref{fig:resultsp_cos}(e, f).

\begin{figure*}[t]
 \centering
 \includegraphics[width=0.7\textwidth]{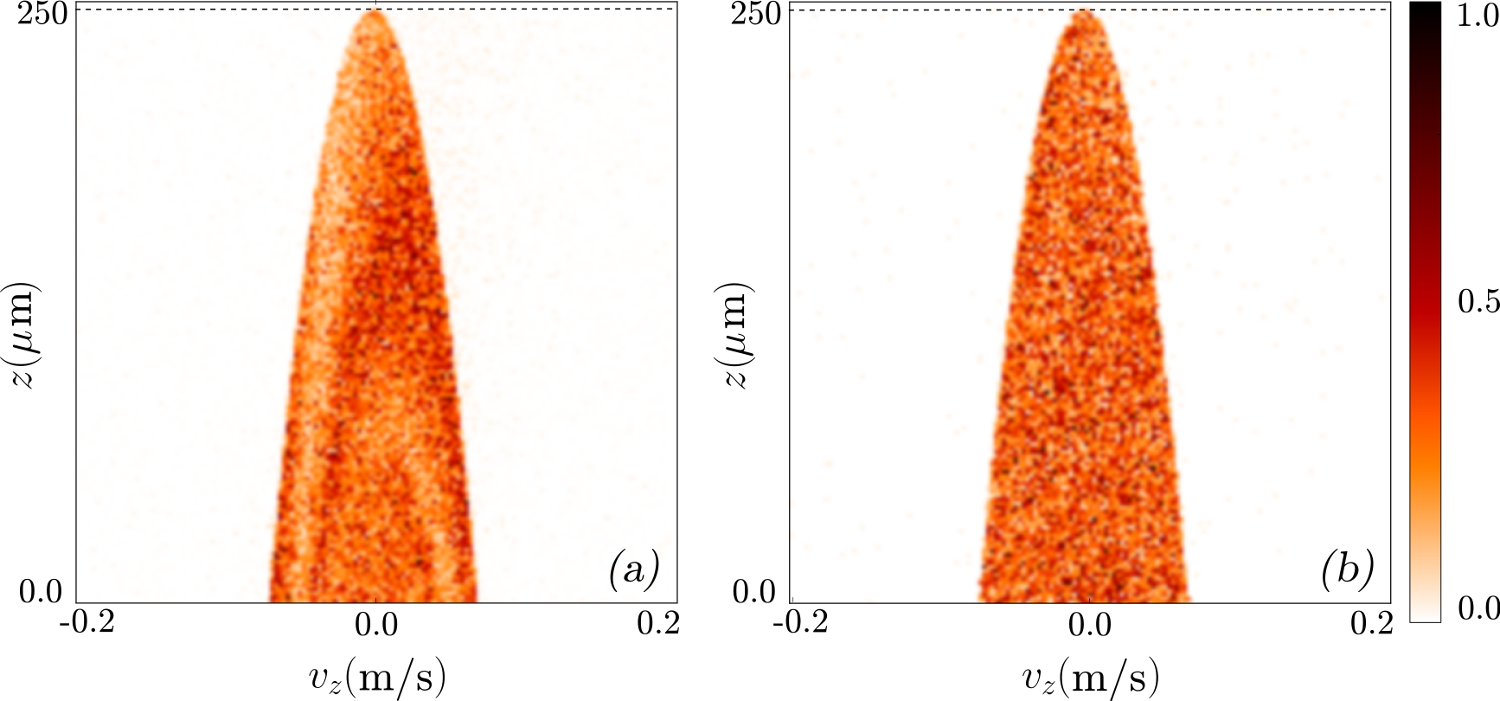}
 \caption{(Color online) Projection of the Poincar\'e section $x = l_x$ onto the ($v_z$, $z$) plane for the filter system with (a) $l_x = 0.1$ m and (b) $l_x = 0.2$ m. Only the regular island is visible since the trajectories in the chaotic sea are spread out over a large volume in the phase space resulting in a much reduced density in the projected surface of section compared to the regular island. The structure in the island (a) is due to nonuniform velocity and position distributions at the ARMS entrance. Due to the large number of bounces, the distribution in the island (b) is more homogeneous.}
 \label{fig:result_PS}
\end{figure*}

\lsdel{Key features for the vectorial velocity selecting and beam shaping properties of the ARMS 
become evident}\ls{Certain key features of the vectorial velocity selection and
beam-shaping properties of the ARMS stand out}: the longitudinal ($v_x$) velocity becomes more monoenergetic reducing the 
initial velocity spread (FWHM) at the collimator opening (see Fig.~\ref{fig:tof}) from 
$\Delta (v_x) \simeq 7 \: m/s$ to $\Delta (v_x) \simeq 2 \: m/s$. The width (HWHM) of 
the angular distribution about the forward direction is reduced to
$\Delta(\theta_z) \simeq 0.3^{\circ}$ while $\Delta(\theta_y) \simeq 7^{\circ}$ 
(see Fig.~\ref{fig:resultsp_cos}(e, f)). This selectivity for the 3D velocity vector 
distinguishes the present filter from the large number of velocity filters in use 
employing chicanes or cranks (see, for example, \cite{Muzychka98, Daum12, Ignatovich90}) which act 
mostly as energy filters selecting neutrons according to their absolute velocity 
$v = \sqrt{v_x^2 + v_y^2 + v_z^2}$. \lsdel{Alternatively,}\ls{The} filters based on quantum interference 
in layered systems (see, for example, \cite{Antonov74, Ignatovich90}) address only the Cartesian 
component perpendicular to layers closely resembling projected band gaps in the 
electronic band structure. In order to \lsdel{inquire}\ls{examine} \lsdel{into}the processes underlying the filter, 
we analyze the Poincar\'e surface of section $x = l_x$ in the phase space projected onto
the $(v_z, z)$ plane for the case of a cosine angular distribution 
(Fig.~\ref{fig:result_PS}(a)) which prominently displays a regular island of 
skipping motion  \lsdel{as}\ls{with} a dense non-uniform distribution. The presence of this regular 
island plays a key role \lsdel{for}\ls{in} the operation of the ARMS. The distribution of the chaotic 
scattering trajectories is barely visible in this plot since they spread out over a 
much larger region in phase space.

\subsection{Analytic estimate of the velocity band-pass}
\label{sec:MCS_an}

Each velocity component distribution can be decomposed into contributions from skipping and chaotic scattering trajectories (blue dotted and green dashed lines in Fig.~\ref{fig:resultsp_cos}). For the skipping part of phase space it is possible to provide an analytical estimate for the velocity range, i.e.\ for the minimum and maximum velocity of \lsdel{neutrons that can pass}\ls{the neutrons passing through} the filter. The shape of the distributions is more difficult to estimate since it strongly depends on the initial velocities in the neutron beam. To obtain the shape of the distributions we thus rely on the CTMC simulations. The velocity limits are, however, independent of the initial distribution (provided it lies within the skipping region of phase space) and can be analytically determined. 

The band-pass in \ls{the} forward direction, $v_x$, determining to a good degree of approximation the energy spread of the UCN beam, can be calculated as follows. The maximum (minimum) velocity corresponds to the minimum (maximum) time of flight between the collimator and the glass mirrors
\begin{equation}
\label{eq:timex}
t_x = l_d / v_x.
\end{equation}

\begin{figure}[b]
 \centering
 \includegraphics[width=0.9\linewidth]{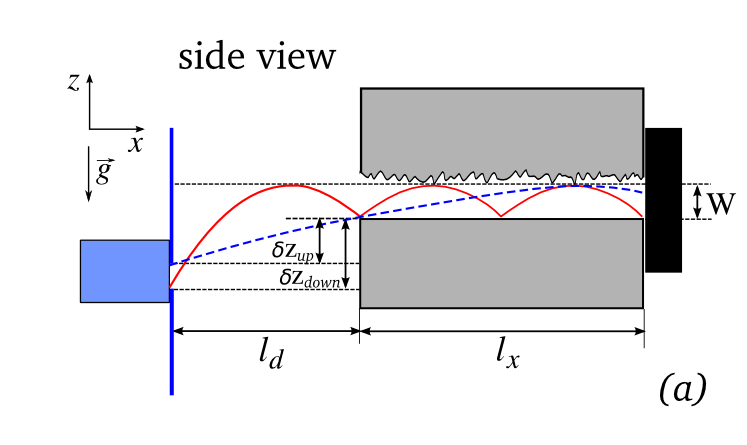}
 \includegraphics[width=0.9\linewidth]{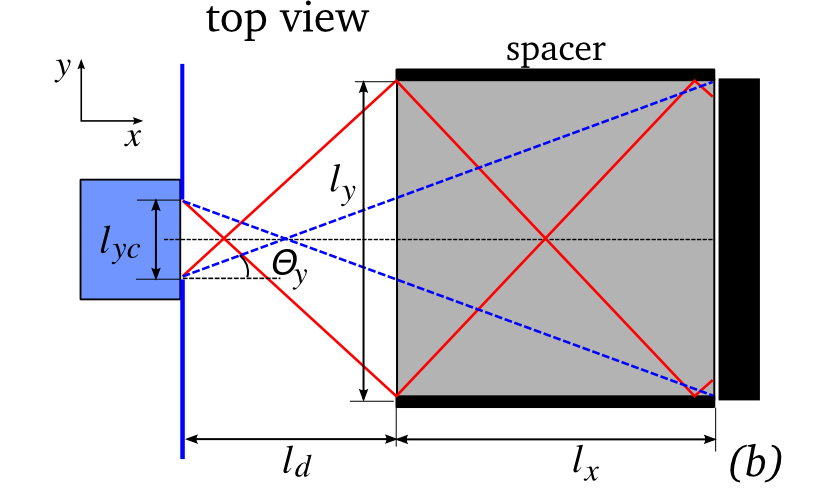}
 \caption{(Color online) Trajectories representing the limiting velocities for transmission. (a) 
 The solid red line corresponds to the trajectory with minimal forward velocity $v_x$, the dashed blue line to the one with maximal $v_x$. (b) 
 The solid red lines correspond to the trajectories launching with the largest angle $\theta_y$ still reflected by the spacers (\lsdel{depicted as black horizontal rectangles}\ls{thick black lines top and bottom}); the dashed lines display the same for the case without spacers.}
 \label{fig:limvel}
\end{figure}

Moreover, neutrons have to enter the mirror system in the skipping region: neutrons do not touch the upper rough surface if the vertical velocity component at the collimator lies in the interval 
\begin{equation}
\label{eq:velzrange}
\sqrt{2g \delta z} \leq v_{z0} \leq \sqrt{2g (\delta z + W)}, 
\end{equation}
where $\delta z$ is the distance between the lower glass plate and an initial $z$-coordinate of the neutron at the collimator opening. \lsdel{Taking into account the motion in $z$-direction allows to}\ls{By taking the motion in $z$-direction into account we can} calculate the neutron time of flight between the collimator and the glass mirrors (Eq.~(\ref{eq:timex})),
\begin{equation}
\label{eq:timeperp}
t_{\perp} = (v_{z0} - v_{zs}) / g,
\end{equation}
where $v_{z0} = v_{z}(t=0)$ is the initial vertical velocity and $v_{zs}$ is the velocity component at the mirror entrance slit,
\begin{equation}
\label{eq:vzs}
v_{zs} = \pm \sqrt{v_{z0}^2 - 2g(\delta z + z_{sl})}.
\end{equation}
Here $z_{sl}$ is the vertical distance from the point where the trajectory enters the ARMS to the lower mirror. Eq.~(\ref{eq:timex}) and Eq.~(\ref{eq:timeperp}) give $v_x$ as a function of $v_{z0}$ and $\delta z$,

\begin{equation}
\label{eq:vx}
v_x = \frac{g l_d}{v_{z0} \pm \sqrt{ v_{z0}^2 - 2 g (\delta z + z_{sl})}}.
\end{equation}
For the fastest (slowest) neutrons the denominator in Eq.~(\ref{eq:vx}) has the smallest (largest) value corresponding to the $-$($+$) sign. Also, the absolute value of $v_{zs}$ in Eq.~(\ref{eq:vzs}) \lsdel{should be  maximal which is reached}\ls{attains its maximum} for $z_{sl}=0$ and \ls{for} $v_{z0}$ at its maximum value \lsdel{obtained from}(Eq.~(\ref{eq:velzrange})). In addition $\delta z$ takes on the smallest (largest) possible value. The two trajectories with maximal and minimal $v_x$ are illustrated in Fig.~\ref{fig:limvel}(a) (side view) and correspond to the velocities
\begin{subequations}
\label{eq:limitvel}
   \begin{equation}
      v_{xmax} = \frac{g l_d}{\sqrt{ 2 g (\delta z_{up} + W)} - \sqrt{2gW}},
   \end{equation}
   \begin{equation}
      v_{xmin} = \frac{g l_d}{\sqrt{ 2 g (\delta z_{down} + W)} + \sqrt{2gW}},
   \end{equation}
\end{subequations}
where $\delta z_{up}$ ($\delta z_{down}$) is the distance between the upper (lower) edge of the collimator opening and the lower glass plate (see Fig.~\ref{fig:limvel}(a)). Eq.~(\ref{eq:limitvel}) represents the analytic estimate for the velocity band-pass for $v_x$. Obviously, the velocity range in Eq.~(\ref{eq:limitvel}) is \lsdel{controlled by geometric parameters of the system only}\ls{only controlled by the geometric parameters of the system} and can be tuned accordingly. 

The range of \lsdel{transmitted}vertical velocity component $v_z$ for \ls{transmitted} trajectories undergoing skipping motion giving the upper bound for the angular divergence in $\theta_z$ can be obtained from the requirement that neutrons do not touch the upper wall inside the mirror slit, i.e.,
\begin{equation}
\label{eq:v_zlim}
|v_z| \leq \sqrt{2gW}. 
\end{equation}
We note that the observed angular divergence $\Delta \theta_z$ (HWHM) is smaller than estimated from the upper bound given by Eq.~(\ref{eq:v_zlim}) (see Fig.~\ref{fig:resultsp_cos}(d, f)).

The range of the $v_y$ component can be found from Fig.~\ref{fig:limvel}(b) representing the top view of the system. \lsdel{Here}\ls{The figure shows} the trajectories with the largest angles that can still be detected\lsdel{are shown}. In the presence of spacers we find 
\begin{subequations}
\label{eq:v_ylim1}
   \begin{equation}
   |v_y| \leq v_{xmax} \frac{l_y + l_{yc}}{2l_d},
   \end{equation}
where $l_{yc}$ is the size of the collimator opening in \ls{the} $y$-direction. Without spacers the allowed velocity range is reduced to
   \begin{equation}
   \label{eq:v_ylim2}
   |v_y| \leq v_{xmax} \frac{l_y + l_{yc}}{2 (l_d + l_x)},
   \end{equation}
\end{subequations}
since \ls{the} neutrons that would have been reflected from the spacers and \lsdel{stay inside}\ls{kept within} the system are now lost. The $v_y$ range \lsdel{depends also only}\ls{is also only dependent} on the geometric parameters of the system and can be tuned as well.

The limits Eq.~(\ref{eq:limitvel} - \ref{eq:v_ylim1}) are depicted in Fig.~\ref{fig:resultsp_cos}(b-d) as dashed vertical lines and coincide with limits for the distributions of the skipping states (blue dotted lines) obtained from Monte Carlo simulations. It should be noted that the band-pass filter limits for the $v_x$ and $v_z$ velocity components are separable, i.e.~independent of the other Cartesian components of the velocity vector and only controlled by geometry. Only the $v_y$ limit depends, in addition to geometric parameters, on the maximum velocity in \ls{the} $x$-direction. Consequently, the ARMS can serve as a band-pass filter for the full vectorial velocity $\vec{v}$, \lsdel{however}\ls{although} only for skipping motion. 

\begin{figure*}[t]
 \centering
 \includegraphics[width=0.9\linewidth]{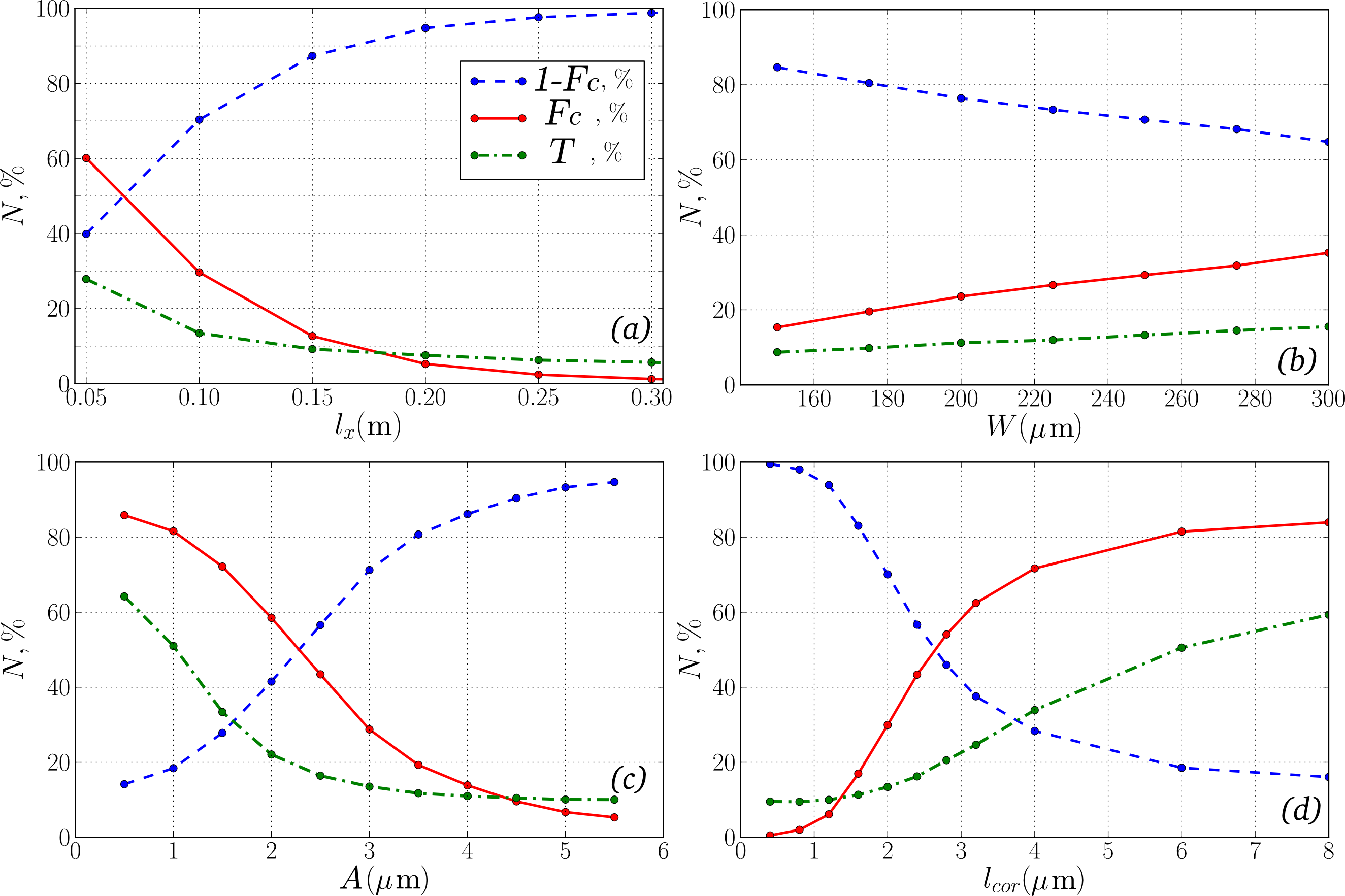}
 \caption{(Color online) Dependence of the fraction of chaotic scattering $F_c$ (solid line), skipping $(1-F_c)$ (dashed line) trajectories and the total transmission probability $T$ through the neutron filter (dotted-dashed line) as a function of (a) the filter length $l_x$; (b) \lsdel{of}the slit width $W$; (c) \lsdel{of}the amplitude $A$ and (d) \lsdel{of}the correlation length $l_{cor}$ of the surface disorder. The parameters of the filter (unless stated otherwise): $W = 250\: \mu$m, $l_x = l_y = 10$ cm, $A = 3\: \mu$m, $l_{cor} = 2\: \mu$m.}
\label{fig:resNL}
\end{figure*}

As the filtering process \lsdel{is operational only}\ls{only functions} for the regular island of skipping motion, it is now crucial to inquire into the relative weight of skipping vs.~chaotic trajectories and to identify parameters to enhance the relative weight of the skipping region of phase space. The fraction of the flux transported by the regular island determines the flux of the \lsdel{beam of UCN}\ls{UCN beam} shaped by the ARMS. For the case depicted in Fig.~\ref{fig:resultsp_cos}, the fraction of chaotic scattering trajectories
\begin{equation}
\label{eq:ch_vol}
F_c = \frac{\Gamma_c}{\Gamma_c + \Gamma_s},
\end{equation}
where $\Gamma_{c,s}$ denotes the volume in phase space pertaining to either chaotic (c) or skipping (s) motion, is $F_c = 0.27$. This fraction can, however, be reduced (see Fig.~\ref{fig:resNL}) by exploiting the fact that chaotic trajectories executing many random bounces \lsdel{at}\ls{off} the mirror are subject to enhanced absorption (see Eq.~(\ref{eq:refl})). For example, for the distributions \lsdel{of}\ls{in} Fig.~\ref{fig:resultsp_cos}, the absorption by wall collisions suppresses the flux of chaotic scattering trajectories at the detector by 12$\%$. In contrast, the flux of neutrons undergoing skipping motion is reduced \lsdel{only by}\ls{by only} 0.2$\%$, because of the drastically different number of collisions for chaotic scattering and skipping motion (for this example $\sim$ 500 and $\sim$ 10, respectively). By further increasing the number of bounces the phase space region associated with chaotic scattering can be \lsdel{even more suppressed}\ls{reduced even further}. Trajectories belonging to $\Gamma_c$ give rise to a broad velocity distribution at the detector extending well beyond $v_{cr}$ and the band-pass (Eq.~(\ref{eq:limitvel}-\ref{eq:v_ylim1}), see Fig.~\ref{fig:resultsp_cos}). 
The low-velocity neutrons correspond to trajectories with a large number of bounces. 
Since the total velocity of these neutrons is less than $v_{cr}$ they do not penetrate 
the walls. The leading loss mechanism is given by Eq.~(\ref{eq:refl}) which reduces 
the number of such neutrons but does not completely remove them. 
A simple estimate for reflection losses (without taking into account the angular dependencies)
is $\sim \exp(- NV_I/V_R)$, where N is number of bounces and $V_I/V_R = 0.0015$. 
Suppressing the amount of these low-velocity neutrons to the $e^{(-3)}$ level would 
require $\sim$ 2000 or more wall collisions or a surface coating of the rough 
surface with a larger absorption cross section. The high-velocity part of the 
spectrum corresponds to neutrons undergoing only a few collisions with the rough wall. 
The normal velocity, $v_{\perp}$, remains \lsdel{still less than}\ls{below} $v_{cr}$ allowing these 
particles to travel through the mirror system without being lost.

\begin{figure*}[p]
 \includegraphics[width=0.9\textwidth]{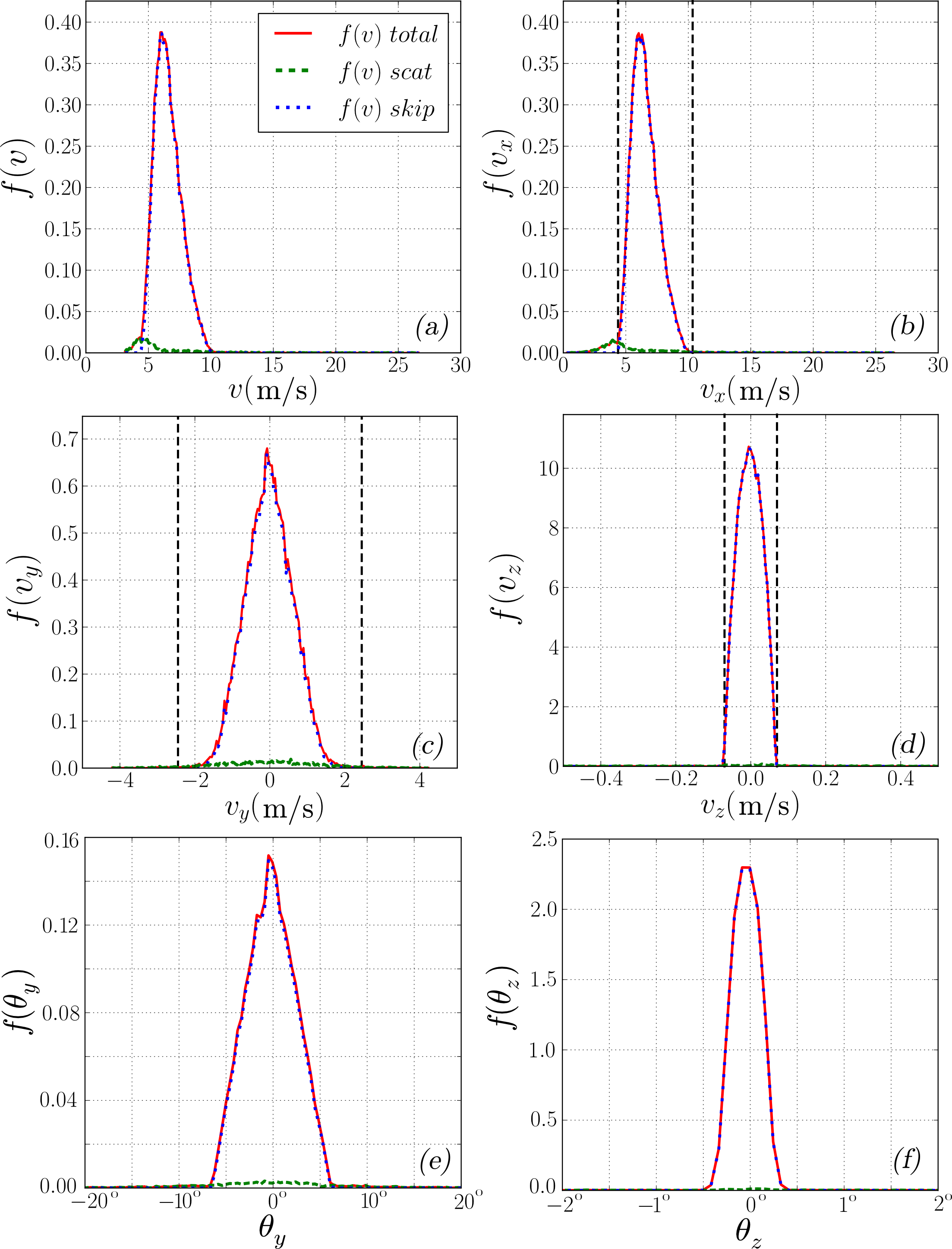}
 \caption{(Color online) Classical trajectory Monte Carlo simulations for the filter as in Fig.~(\ref{fig:resultsp_cos}), \lsdel{however with doubled length}\ls{however with twice the length} $l_x = 20 \: cm$ and without spacers. Distributions of: (a) the absolute value of the total velocity $f(v)$; (b-d) the velocity components $f(v_x)$, $f(v_y)$ and $f(v_z)$; (e, f) angular distributions $f(\theta_y)$ and $f(\theta_z)$ (solid line). Contributions from gravitational skipping and chaotic scattering trajectories are indicated by dotted blue and dashed green lines, respectively. Vertical lines correspond to analytical results for the velocity band-pass (see Eq.(\ref{eq:limitvel}-\ref{eq:v_ylim1})). The parameters of the filter: $W = 250\: \mu$m, $l_y = 10$ cm, $A = 3\: \mu$m, $l_{cor} = 2\: \mu$m.}
\label{fig:resultnosp_cos}
\end{figure*}

One more aspect to be discussed is the influence of spacers. Our Monte Carlo simulations indicate that 34$\%$ of the total number of trajectories (24$\%$ from skipping motion and 10$\%$ from chaotic motion)  are reflected from the spacers. We consider spacers to be flat and \lsdel{to be}perfect specular reflectors, i.e.\ only $v_y$ changes sign after scattering. Realistically, the spacers do have a rough surface as well as other surface imperfections and thus may also give rise to randomization of the reflection angle. This can \lsdel{lead to an additional contribution}\ls{contribute} to the irregular part of \ls{the} phase space. For example, skipping trajectories can become chaotic scattering trajectories after a single reflection from a spacer. In the following we focus on the system without spacers.

\subsection{Suppressing the chaotic scattering contribution}
\label{sec:MCS_mod}
{}
Controlling chaotic scattering trajectories requires \lsdel{the}tuning \lsdel{of}the surface disorder and the length of the mirror system. In particular, the chaotic part of the phase space $F_c$ can be suppressed by increasing the number of wall collisions as discussed in section \ref{sec:MCS_an}. Enhancing the number of bounces is accomplished by either increasing the length $l_x$ or decreasing the width $W$. Furthermore, a variation of the amplitude $A$ and correlation length $l_{cor}$ allows \ls{us} to modify the number of wall collisions and, moreover, to control scattering into states with $v_{\perp} > v_{cr}$. 

The sensitivity to \lsdel{the variation of}\ls{variations in} the available control parameters is shown in Fig.~\ref{fig:resNL}. While the fraction of chaotic phase space $F_c$ strongly decreases with $l_x$, the variation with $W$ is much weaker. Accordingly, $F_c$ does not simply scale with $W/l_x$. This is, in fact, due to the presence of additional relevant length scales, $A$ and $l_{cor}$. $F_c$ and, thus, the chaotic scattering contribution outside of the filter band-pass can be minimized for a long mirror system (large $l_x$) with short-ranged surface disorder (small $l_{cor}$) and large disorder amplitude $A$. Clearly, the price to pay for this purification of the beam is the reduction of the overall transmission probability (typically $\lesssim$ 10 $\%$).

\begin{figure}[t]
 \centering
 \includegraphics[width=0.5\textwidth]{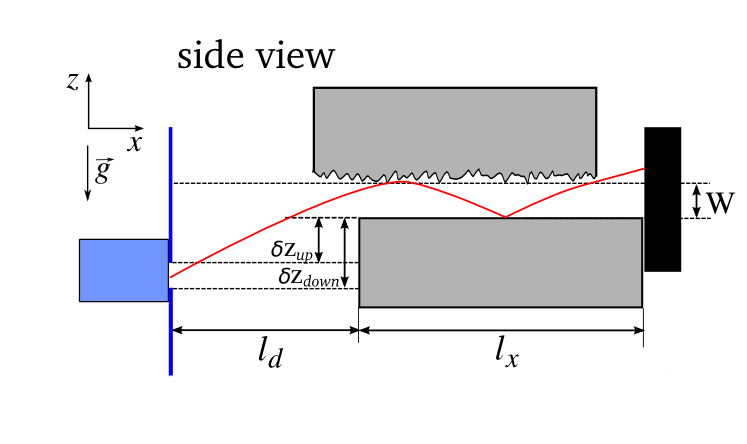}
 \caption{(Color online) The side view of the experimental setup.}
 \label{fig:xz_exp}
\end{figure}

\begin{figure*}[b]
 \centering
 \includegraphics[width=1.0\linewidth]{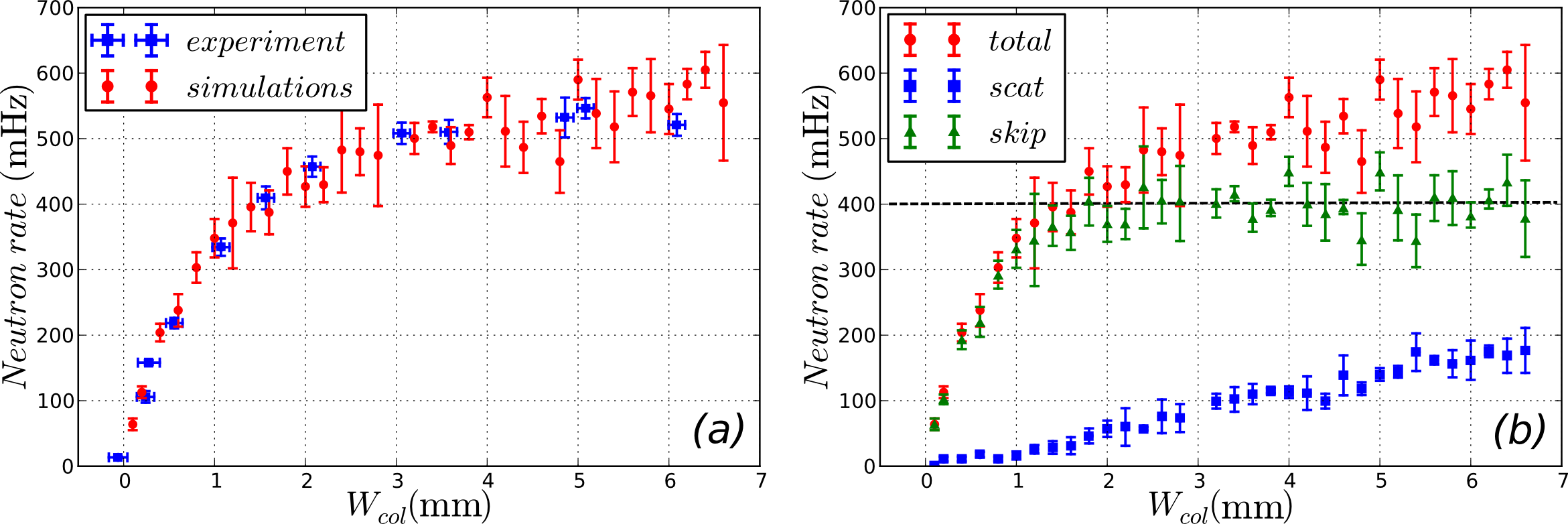}
 \caption{(Color online) The neutron transmission rate through the ARMS as a function of the collimator opening $W_{col}$. (a) The experimental (blue squares) and calculated (red circles) data. (b) The calculated data (red circles) and its decomposition into chaotic scattering (blue squares) and gravitational skipping (green triangles) contributions. The horizontal dashed line shows a saturation of the amount of gravitational skipping trajectories.}
 \label{fig:exp1}
\end{figure*}

Since a variation of the amplitude or the correlation length of the surface disorder might be difficult to achieve experimentally, a practical approach is to vary the length of the mirrors. In Fig.~\ref{fig:resultnosp_cos} we present distributions of the total velocity and its Cartesian components $v_x$, $v_y$ and $v_z$ in the absence of the spacers and for \lsdel{the doubled}\ls{twice the} length ($l_x = 20$ cm) of the ARMS compared to the previous settings (Fig.~\ref{fig:resultsp_cos}). The initial angular distributions entering the CTMC simulation are, as before, cosine distributions (results for uniform distributions are hardly different). The contribution from chaotic scattering trajectories to the velocity distributions is now suppressed and the spectra consist predominantly of skipping trajectories (see Fig.~\ref{fig:resultsp_cos} for comparison). Although the high-velocity part of the spectra (see Fig.~\ref{fig:resultnosp_cos}(a, b)) outside the upper filter limit has disappeared, there is still a small contribution of chaotic scattering trajectories at velocities below the lower cut-off of the filter. This part corresponds to neutrons with an energy smaller than the Fermi potential of the mirrors and with a number of bounces not high enough to be absorbed due to repeated wall collisions. Because of an increasing number of bounces the distribution within the regular island at the detector position, shown in Fig.~\ref{fig:result_PS}(b), \lsdel{becomes}\ls{is} more homogeneous \lsdel{compared to}\ls{than that of} Fig.~\ref{fig:result_PS}(a). From the fraction of the total number of detected neutrons (Fig.~\ref{fig:resNL}(a)), it can be seen that the accumulation time for good statistics is of the same order as for the setup previously discussed in section \ref{sec:MCS_geo}.

\section{First experimental test}
\label{sec:experiment}

\lsdel{We have performed the first experimental test of the beam shaper properties of the ARMS at the ILL (Institute Laue-Langevin). Due to the limited UCN flux, a direct time of flight measurement of the resulting velocity distribution is not possible. That is why we probe the reduction of the $\theta_z$ angular divergence by the confinement of the trajectories to the skipping island.}\ls{We performed the first experimental test of the ARMS' beam-shaping properties at the ILL (Institut Laue-Langevin). Due to the limited UCN flux, it was not possible to make direct time-of-flight measurements of the resulting velocity distribution. We therefore probed the reduction of the  angular divergence by confining the trajectories to the skipping island. The ARMS' geometry was slightly modified, by displacing the upper mirror horizontally by $(1.5 \pm 0.1)$ cm (see Fig.~\ref{fig:xz_exp}).} \lsgreen{This shift was introduced to be able to align the mirrors properly in the experiment and is fully accounted for in the corresponding (CTMC) calculations. Differences between the results for mirror setups with and without this shift will be discussed below.} \lsdel{The other geometric parameters of the filter was chosen}\ls{For the filter's other geometric parameters, we chose} $l_x = (20 \pm 0.1)$ cm, $l_y = (15 \pm 0.1)$ cm, $W = (193 \pm 1)\: \mu$m, $l_d = (15.5 \pm 0.1)$ cm, $A = (3.0 \pm 0.75) \: \mu$m, $l_{cor} = (2.0 \pm 0.75) \: \mu$m.

\lsdel{In a first step}\ls{In the first part of the test} we measured the dependence of the transmitted neutron flux through the filter as a function of collimator opening $W_{col}$. The position of the upper collimator plate was fixed throughout the experiment. We set the position of the upper collimator plate at a distance to the lower mirror of $(2.5 \pm 0.08)$ mm. The collimator opening was varied by moving the lower collimator plate. \lsdel{By changing $W_{col}$ the flux entering the ARMS can be modified}\ls{The flux entering the ARMS can be modified by changing $W_{col}$}. The measured outgoing flux (Fig.~\ref{fig:exp1}(a)) is controlled by the fraction of incident flux resulting in the regular island of skipping motion. In agreement with the simulation, the outgoing flux saturates for $W_{col} \gtrsim 2 \: mm$ (see Fig.~\ref{fig:exp1}(b)) indicating that the population of skipping trajectories cannot be enlarged by further increasing $W_{col}$. \lsdel{For the simulation we used the initial velocity distribution similar to the one from Fig.~\ref{fig:tof}, but with the maximum shifted by $2$ m/s to account for different position of the beam pipe compare to the TOF measurements of 2008.}\ls{For the simulation we used an initial velocity distribution similar to that of Fig.~\ref{fig:tof}; the maximum was shifted by $2$ m/s however, to account for the difference in the position of the beam pipe compare to the TOF measurements in \cite{Jenke11a}.}

\begin{figure*}[t]
 \centering
 \includegraphics[width=1.0\linewidth]{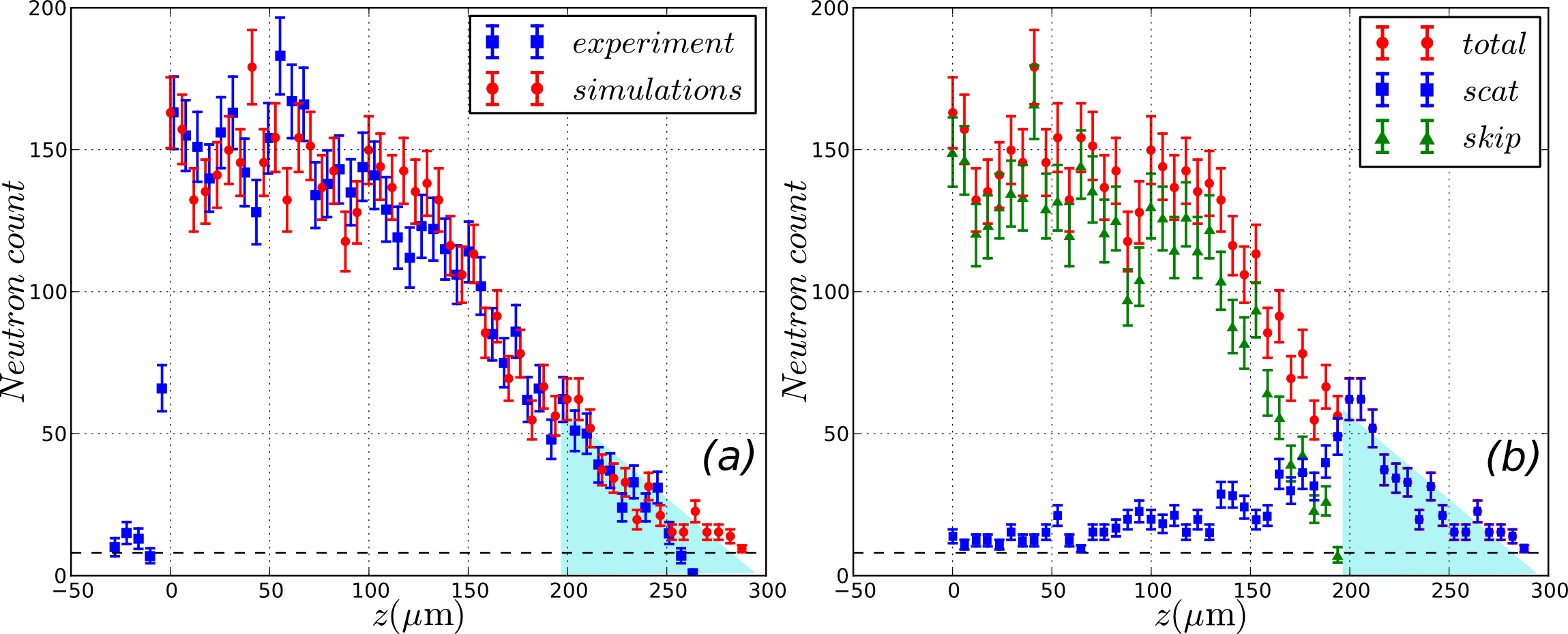}
 \caption{(Color online) The $z$-coordinate distribution of the detected neutrons measured with a track detector. (a) The experimental (blue squares) and calculated (red circles) data. (b) The calculated data (red circles) and its decomposition into chaotic scattering (blue squares) and gravitational skipping (green triangles) contributions. A horizontal dashed line corresponds to the background noise in the experiment. The light blue (light gray) area shows the contribution of neutrons with predominantly chaotic scattering trajectories.}
 \label{fig:exp2}
\end{figure*}

In the second \lsdel{step}\ls{part of the test}, we measured the position distribution $P(z)$ with the track detector {\cite{Jenke13}}. \lsdel{Trajectories that belong to the regular island can reach only position}\ls{The trajectories belonging to the regular island can only reach positions} $z \leq W$. \lsgreen{By contrast, trajectories belonging to the chaotic sea that would hit the rough upper surface if the upper mirror had not been displaced horizontally (see Fig.~\ref{fig:xz_exp}) are now able to reach higher values of $z \geq W$ because of the slit opened up by the shift.} The $P(z)$ distribution allows \ls{us} to determine the flux associated with the regular island that is both confined in space ($z \leq W$) and angle $\theta_z \lesssim \Delta \theta_z$. For this measurement, we fixed the position of the upper and lower collimator plates relative to the lower filter mirror at $(3.3 \pm 0.1)$ mm and $(7.3 \pm 0.1)$ mm \lsdel{respectively and put the track detector after the filter which registered the position ($y$- and $z$-coordinates) of neutrons having passed through the ARMS}\ls{respectively. We placed the track detector recording the position ($y$- and $z$-coordinates) of the neutrons which had traversed the ARMS after the filter.}
The $z$-coordinate distribution of the detected neutrons after integration over the $y$-coordinate is shown in Fig.~\ref{fig:exp2}\ls{,} together with the result of the simulation.  The background noise level is indicated by the horizontal dashed line. \lsdel{To a good degree of approximation $P(z)$ is confined to the distribution associated with skipping motion}\ls{$P(z)$ is confined to the distribution associated with the skipping motion to a good degree of approximation} (Fig.~\ref{fig:exp2}(b)). However a small fraction of $(4.45 \pm 0.73)\%$ lies above $z=W$ and is associated with chaotic trajectories. \lsdel{This is a consequence of the fact that the position of the collimator plates was chosen to generate high neutron flux through the ARMS (due to time limitation of the experiment) which was not optimized for the filter.}\ls{This is because the position of the collimator plates  was not optimized for the filter, as the position had been chosen in order to ensure high neutron flux through the ARMS (given the time limits of the experiment).} Therefore the velocity filter allowed for the transmission of a small amount of neutrons on chaotic scattering trajectories. This fraction can be easily suppressed \lsdel{by choosing an improved setting of the collimator slits}\ls{by adjusting the collimator slit settings}.

This first test demonstrates that a well collimated beam of UCN can be formed by transmission through the ARMS. The $P(z)$ distribution provides not only direct evidence for the spatial confinement but also indirect evidence for the small angular spread in $\theta_z$. 

\section{Conclusion}
\label{sec:conc}
\lsdel{In the present work, we demonstrate}\ls{We demonstrate in this work} that an absorbing-reflecting mirror system (ARMS) can be used as \ls{a} band-pass filter for the 3D velocity vector $\vec{v}$ of ultracold neutrons. Full three-dimensional Monte Carlo simulations of the propagation of UCN through the system have been performed. The efficiency of the filter is explored as a function of the geometric control parameters\ls{,} and the underlying scattering phase space structure is analyzed. 

The proposed setup features a mixed phase space, corresponding to two types of motion: regular skipping trajectories bouncing off the lower wall only\ls{,} and chaotic scattering trajectories which \ls{also} scatter \lsdel{also}off the upper plate with surface disorder. These different classes of motion give different contributions to the total velocity distribution of the detected neutrons. The fraction of neutrons contributing to the transmitted flux in the irregular part of the phase space is controlled by the number of scattering events at the rough absorbing mirror\lsdel{and}\ls{; they} can thus be effectively suppressed by increasing the length of the mirror system or by increasing the roughness. We have shown that by optimizing the geometric parameters of the filter or the surface disorder, the fraction of neutrons arriving at the detector originating from the chaotic sea of the phase space can be reduced to $5 \% $ (or less) of the total number of the detected neutrons. 
Since skipping motion is regular and well controlled, \lsdel{the transmission is allowed only for a limited window of the full vectorial velocities giving rise to a band-pass filter}\ls{transmission is only possible for a limited window of full vectorial velocities, thus creating a band-pass filter}. \lsdel{Its limits}\ls{The filter's limits} can be determined analytically. They depend only on the geometric parameters of the system and are independent of the initial velocity distributions at the exit of the neutron guide\ls{,} which are usually only partly known. \lsdel{The proposed system delivers the neutrons coming mostly from the regular island with a well-defined velocity range}\ls{The system proposed delivers neutrons emanating essentially from the regular island with a well-defined velocity range,} while suppressing the trajectories from the irregular part of the phase space. \lsdel{Consequently,}\ls{We conclude that} the ARMS can serve as a shaper for beams of UCN with small angular divergence.


\lsgreen{We expect this type of filter system to be highly beneficial for a variety of experiments with ultracold neutrons. In particular it could be used as a convenient preparation stage for sampling a neutron beam of a desired velocity distribution. For example, beams with a well-controlled transverse velocity are required for experiments probing scattering dynamics in the deep quantum regime (see, e.g., \cite{Feist06}) where quantum states in the gravitational field can be explored.}
\section*{Acknowledgments}
We gratefully acknowledge support from the doctoral colleges CMS (TU-Vienna) and Solids4Fun (FWF), from ViCom (SFB 041-ViCom), from the Austrian Science Fund (FWF) under Contract No. I529-N20 and No.I862-N20 and the German Research Foundation (DFG) as part of the Priority Programme (SPP) 1491 "Precision experiments in particle and astrophysics with cold and ultracold neutrons". We also gratefully acknowledge support from the French L'Agence nationale de la recherche ANR under contract number ANR-2011-ISO4-007-02, Programme Blanc International - SIMI4-Physique. We thank K.~Mitsch for the optical read-out of the track detector and T.~Brenner for technical assistance.

\end{document}